\documentclass[preprint,10pt]{elsarticle}

\usepackage{amssymb,amsmath}
\usepackage{longtable}
 \usepackage{rotating}
 \usepackage{multirow}
 \usepackage{booktabs}

\newtheorem{remark}{Remark}
\newtheorem{theorem}{Therorem}
\newtheorem{lemma}{Lemma}

\newtheorem{example}{Example}

\journal{AAECC}

\begin{document}

\begin{frontmatter}



\title{A further study on the linear complexity of new binary cyclotomic sequence of length $p^r$}
\author[abc,rvt]{Zhifan Ye}
\ead{yzffjnu@163.com}
\author[rvt]{Pinhui Ke\corref{cor}}
\ead{keph@fjnu.edu.cn}
\cortext[cor]{Corresponding author}
\author[els]{Chenhuang Wu}
\ead{wuchenhuang2008@126.com}

\address[abc]{Department of Mathematics and Physics, Fujian Jiangxia University, Fuzhou, Fujian, 350108, P. R. China}
\address[rvt]{Fujian Provincial Key Laboratory of Network Security and Cryptology, Fujian Normal University, Fuzhou, Fujian, 350117, P. R. China}
\address[els]{School of Mathematics, Putian University, Putian, Fujian, 351100, P. R. China}

\begin{abstract}
Recently, a conjecture on the linear complexity of a new class of generalized cyclotomic binary sequences of period $p^r$ was proposed by Z. Xiao et al. (Des. Codes Cryptogr., DOI 10.1007/s10623-017-0408-7). Later, for the case $f$ being the form $2^r$ with $r\ge 1$, Vladimir Edemskiy proved the conjecture (arXiv:1712.03947). In this paper,  under the assumption of $2^{p-1} \not\equiv 1 \bmod p^2$ and $\gcd(\frac{p-1}{{\rm {ord}}_{p}(2)},f)=1$, the conjecture proposed by Z. Xiao et al. is proved for a general $f$ by using the Euler quotient. Actually, a generic construction of $p^r$-periodic binary sequence based on the generalized cyclotomy  is introduced in this paper, which admits a flexible support set and includes Xiao's construction as a special case, and then an efficient method to compute the linear complexity of the sequence by the generic construction is presented, based on which the conjecture proposed by Z. Xiao et al. could be easily proved under the aforementioned assumption.

\end{abstract}

\end{frontmatter}

\section{Introduction}

Sequences with good pseudo-random properties are widely used in simulation, ranging systems, code-division multiple-access
systems, and cryptography \cite{V,Gol}.  Linear complexity is one of the most important properties of a sequence for its applications in cryptography. The linear
 complexity $L((s_{i}))$ of an $N$-periodic sequence $(s_{i})$ over a finite field $\mathbb F_{q}$ is the smallest positive integer
 $L$ for which there exist constants $c_{0},c_{1},\ldots,c_{L-1}\in \mathbb F_{q}$ such that
 $s_{i+L}=c_{L-1}s_{i+L-1}+\cdots+c_{1}s_{i+1}+c_{0}s_{i}$, for all $i\geq 0$ \cite{Gol}. By the Berlekamp-Massey algorithm, for a
 sequence with period $N$, if its linear complexity is large than $\frac{N}{2}$, then it is considered good with respect to the linear
 complexity \cite{Massey}.

Blahut's theorem is a common method to compute the linear complexity of a sequence. Let $(s_{i})$ be an $N$-periodic sequence,
then the generating polynomial of the sequence $(s_{i})$ is defined as
$$S(x)=s_{0}+s_{1}x+\cdots+s_{N-1}x^{N-1}.$$
Let $\alpha$ be a primitive $N$-th root of unity in an extension field of $\mathbb F_{q}$. Then by Blahut's theorem, the linear
complexity of sequence $(s_{i})$ is equal to
$$L((s_{i}))=N-|\{t|S(\alpha^{t})=0,t=0,1,\ldots,N-1\}|.$$
 Thus, one could determine the linear complexity of a sequence $S$ by counting the number of roots of $S(x)$ in the set
 $\{\alpha^{t}|t=0,1,\ldots,N-1\}$.

For an odd prime $p$, let $p=ef+1$ and $g$ be a primitive root modulo $p^2$. It is well known that $g$ is also a primitive root modulo $p^j$ for each integer $j\ge 1$. For an integer $j\ge 1$, define
$$D_{n}^{(p^{j})}=\{g^{n+kfp^{j-1}}\bmod p^{j}:0\leq k <e-1\},$$
where $n=0,1,\ldots,p^{j-1}f-1$. Then we get generalized cyclotomic classes of order $p^{j-1}f$ with respect to $p^{j}$. It was shown that $\{D_0^{(p^{j})}, D_1^{(p^{j})}, \ldots, D_{p^{j-1}f-1}^{(p^{j})}\}$ forms a partition of $\mathbb{Z}_{p^j}^*$ in \cite{Zheng}.

Recently, when $f=2^r$ with $r\ge 1$, a new family of binary sequences $(s_i)$ of length $p^2$ were introduced by Xiao et al. \cite{xiao} via the aforementioned cyclotomic classes for the cases of $j=1$ and 2, where
\begin{equation*}
s_{i}=\left\{
                                    \begin{array}{ll}
                                      1, & \mbox{if}~ i\in \displaystyle  \bigcup_{j=0}^{\frac{f}{2}-1}pD_{j}^{(p)}\bigcup_{j=0}^{\frac{pf}{2}-1}D_{j}^{(p^{2})}\bigcup\{0\} ;\\
                                      \\
                                      0, & \mbox{otherwise} .
                                    \end{array}
                                  \right.
\end{equation*}
In \cite{xiao}, Xiao et al. determined the linear complexity of the sequences and showed that the sequences have large linear complexity if $p$ is a non-Wieferich prime.

As it was remarked at the end of Ref. \cite{xiao}, it is natural to generalize above construction of new generalized cyclomomic sequences of period $p^2$ to the case of period $p^m$ for an odd prime $p$ and an integer $m\ge 3$ as follows. For $0\le i\le p^m-1$, let $b$ be an integer with $0\le b\le p^{m-1}f-1$, define
\begin{equation}\label{eqn1}
s_{i}=\left\{
                                    \begin{array}{ll}
                                      1, & \mbox{if}~ i\in C_1 ;\\
                                      \\
                                      0, & \mbox{otherwise} ,
                                    \end{array}
                                  \right.
\end{equation}
where
$$C_1=\displaystyle  \bigcup_{j=1}^{m}\bigcup_{i=0}^{\frac{p^{j-1}f}{2}-1}p^{m-j}D_{(i+b)\pmod {p^{j-1}f}}^{(p^j)}\cup \{0\}.$$
For above sequence, the authors failed to determine the linear complexity by the similar method in \cite{xiao} and proposed a conjecture as follows.

{\bf Conjecture:} Let $p$ be a non-Wieferich odd prime and let $(s_{i})$ be a generalized cyclotomic binary sequence of period $p^r$ defined by (\ref{eqn1}). Then its linear complexity is given by
\begin{equation*}
L((s_i))=\left\{
                                    \begin{array}{ll}
                                      p^r-\frac{p-1}{2}-\delta(\frac{p^r+1}{2}), & 2\in D_0^{(p)} ;\\
                                      \\
                                      p^r-\delta(\frac{p^r+1}{2}), & 2\notin D_0^{(p)},
                                    \end{array}
                                  \right.
\end{equation*}
where $\delta(t)=1$ if $t$ is even and $\delta(t)=0$ if $t$ is odd.

In \cite{Edemskiy}, Vladimir Edemskiy proved the conjecture for the case $f$ being the form $2^r$ with $r\le 1$. In this paper, we will further study this problem. Different from the method in \cite{Edemskiy}, we will introduce a generic construction of generalized cyclotomic binary sequence of period $p^n$, in which the generalized cyclotomic classes could be chosen arbitrary as their support sets. Thus, our generic construction includes the sequence (\ref{eqn1}) as a special case. Then for the case $2^{p-1} \not\equiv 1 \bmod p^2$ and $\gcd(\frac{p-1}{{\rm{ord}}_{p}(2)},f)=1$, we will give an efficient method to determine the linear complexity over $\mathbb F_{2}$ of the generalized cyclotomic binary sequence derived from the generic construction by using Euler quotient. As a byproduct, for the case $\gcd(\frac{p-1}{{\rm{ord}}_{p}(2)},f)=1$, the conjecture given by Z. Xiao et al. is proved.

The remainder of this paper is organized as follows. In Section 2, we will first introduce a generic construction of generalized cyclotomic binary sequence of period $p^r$, which includes the construction in \cite{xiao} as a special case. Then, we will present our main result. The proof of the main result will be given in Section 3. Firstly, we will recall the definition of Euler quotients and its cyclotomic characterization. We will also establish the relationship between the generalized cylcotomic classes derived from Euler quotients and the generalized cylcotomic classes defined in \cite{Zheng}. Then some useful lemmas will be given. Lastly, by using the Euler quotient, the main result will be proved. Some examples will also be illustrated.

\section{A generic construction and the main result}


\subsection{A generic construction }

Let $p$ be an odd prime with  $p=ef+1$ and $r\ge 1$ be an integer. Denote $Y_{t}$ a subset of $\mathbb{Z}_{p^{t-1}}\times \mathbb{Z}_{f}$, i.e. $Y_{t}\subseteq \{(l,m):l\in\mathbb{Z}_{p^{t-1}},m\in\mathbb{Z}_{f}\}$, for $t=1,\ldots,r$. Define $X_{t}=\{l+mp^{t-1}:(l,m)\in Y_{t}\}$, and
$$C_{1}=\displaystyle  \bigcup_{i\in X_{1}}p^{r-1}D_{i}^{(p)}\bigcup_{i\in X_{2}}p^{r-2}D_{i}^{(p^{2})}\bigcup\cdots\bigcup_{i\in X_{r}}D_{i}^{(p^r)}\bigcup\{0\}.$$
Then a $p^r$-periodic sequence can be defined as follows
\begin{equation}{\label{newseq}}
s_{i}=\left\{
                                    \begin{array}{ll}
                                      1, & \mbox{if}~ i\in C_{1};\\
                                      \\
                                      0, & \mbox{otherwise} .
                                    \end{array}
                                  \right.
\end{equation}

\begin{remark} Choosing $Y_{t}=\mathbb{Z}_{p^{t-1}}\times \mathbb{Z}_{\frac{f}{2}}~for~ t=1,\ldots,r$, the sequence in $(\ref{newseq})$ is just the sequence defined in $(\ref{eqn1})$. Thus the sequence defined in \cite{xiao} is a special case of the sequence in our generic construction. In the sequel, we denote $D_{l+mp^{j-1}}^{(p^{j})}$ by $D_{(l,m)}^{(p^{j})}$ for convenience.
\end{remark}

Define ${\rm{ord}}_{p}(2)$ the order of $2$ modulo $p$, since the linear complexity of sequence defined in (2) is depended on ${\rm{ord}}_{p}(2)$, support set $X$ and $f$, let us compare our result with the result given in \cite{Edemskiy} in the following table.

\begin{table}[!htbp]
\centering
\caption{Difference between the result in this paper and the result in \cite{Edemskiy}}\label{tab:aStrangeTable}
\begin{tabular}{ccc}
\toprule
& Ref.\cite{Edemskiy} & This paper\\
\midrule
limitation of ${\rm{ord}}_{p}(2)$& no & $\gcd(\frac{p-1}{{\rm{ord}}_{p}(2)},f)=1$\\
selection of $Y_{t}$ & $\mathbb{Z}_{p^{t-1}}\times \mathbb{Z}_{\frac{f}{2}}$& no limitation\\
limitation of $f=\frac{p-1}{e}$& $2^{r}$& no\\
\bottomrule
\end{tabular}
\end{table}

\subsection{Main result}

For the sequence defined in (\ref{eqn1}), if $f$ has the form of a power of 2, its linear complexity over $\mathbb{F}_2$ was
 determined under the assumption of $2^{p-1}\not\equiv 1 \bmod p^{2}$ in \cite{xiao,Edemskiy}. However, for the linear complexity of the sequence in (\ref{newseq}), the situation will be very different, since in our generic construction, the support sets can be chosen arbitrary and $f$ can be any positive divisor of $p-1$, without the limitation of being a power of 2.
Following is our main theorem of this paper.
\begin{theorem} {(\bf Main theorem)}
Let the sequence $S$ be defined as $(2)$. If $2^{p-1} \not\equiv 1 \bmod p^2$ and $\gcd(\frac{p-1}{{\rm ord}_{p}(2)},f)=1$, then the linear complexity of $S$ over $\mathbb{F}_2$, can be represented as
$$L((S))=\delta+(p-1)\sum_{i=0}^{r-1}\delta_{i}p^{i},$$
where

$$
\delta=\left\{
                                    \begin{array}{ll}
                                      1, & \mbox{if}~ 2|\displaystyle \sum_{j=1}^{r}e|X_{j}| ;\\
                                      0, & \mbox{otherwise}.
                                    \end{array}
                                  \right.
$$
and $\delta_{i}\in \{ 0,1 \}$, $i=0,\ldots,r-1$, which can be determined by Lemma 6.
\end{theorem}

\begin{remark}
As we have remarked in Remark 1, the sequence defined in $(\ref{eqn1})$ can be constructed by choosing  $Y_{t}=\mathbb{Z}_{p^{t-1}}\times \mathbb{Z}_{\frac{f}{2}}~for~ t=1,\ldots, r$. In this case,  $\delta_{i}$, $i=0,\ldots,r-1$ can be easily verified to be 1 by Lemma 6. Furthermore, by definition, we have $$|X_{j}|=\frac{fp^{j-1}}{2}, \mbox{and}~~ \sum_{j=1}^{r}e|X_{j}|=\frac{p^r-1}{2}.$$ Thus,
$$L((s))=\delta+(p-1)\sum_{i=0}^{r-1}\delta_{i}p^{i}=\delta(\frac{p^r-1}{2})+p^r-1=p^r-\delta(\frac{p^r+1}{2}).$$
Hence, under the assumption of $2^{p-1} \not\equiv 1 \bmod p^2$ and $\gcd(\frac{p-1}{{\rm ord}_{p}(2)},f)=1$, the conjecture is proved.
\end{remark}

\section{Proof of the Main Theorem }
In this section, we will prove the main theorem given in Section 2.2. To this end, we will recall the definition of Euler quotient and  will give some auxiliary lemmas firstly, based on which the main theorem can be easily proved. Two examples will also be given to illustrate the efficiency of our method.

\subsection{Euler quotient and its cyclotomic characterization}

For an odd prime $p$, an integer $r\geq 1$ and $u$ with $\gcd(u,p)=1$, the Euler quotient $Q_{p^r}(u)$ modulo $p^{r}$ is defined as
the unique integer with
$$Q_{p^r}(u)\equiv \frac{u^{\phi(p^{r})}-1}{p^{r}}\bmod p^{r}, ~~~~~0\leq Q_{r}(u)\leq p^{r}-1,$$
where $\phi(\cdot)$ is the Euler function and it is also defined that
$$Q_{p^r}(kp)=0,~~~~~~k\in \mathbb Z.$$
The following properties are useful for the calculation of Euler quotient,
$$
Q_{p^r}(u +kp^{r})\equiv Q_{r}(u)-kp^{r-1}u^{-1}(\bmod p^{r});
$$
$$
Q_{p^r}(uv)\equiv Q_{r}(u) + Q_{r}(v) (\bmod p^{r});
$$
for $\gcd(u,p) = 1$ and $\gcd(v,p) = 1$.
A partition of $\mathbb{Z}_{p^{r+1}}^*$ is then induced by Euler quotient. In detail, let
$$\widetilde{D}_l^{(p^{r+1})}=\{u:0\leq u \leq p^{r+1}-1, \gcd(u,p)=1, Q_{p^r}(u)=l\}, $$
where $0\leq l\leq p^r-1$. Then, $\mathbb{Z}_{p^{r+1}}^*=\bigcup_{l=0}^{p^r-1}\widetilde{D}_l^{(p^{r+1})}$.
In \cite{Chen5}, it is shown that $\widetilde{D}_l^{(p^{r+1})}$  can also be represented as
$$\widetilde{D}_l^{(p^{r+1})}=\{g^{l+kp^{r}}\bmod p^{r+1}:0\leq k <p-1\}, $$
where $g$ is a primitive element modulo $p^{r+1}$ with $Q_{r}(g)=1$. Meanwhile, $\widetilde{D}_l^{(p^{r+1})}$ are also called generalized cyclotomic classes of order $p^r$ with respect to $p^{r+1}$.

\begin{remark}Comparing with the generalized cyclotomic classes defined in \cite{Zheng, xiao}
$$D_{n}^{(p^{j})}=\{g^{n+kfp^{j-1}}\bmod p^{j}:0\leq k <e-1\},$$
where $n=0,1,\ldots,p^{j-1}f-1$, it is easy to see that
$$\widetilde{D}_{l}^{(p^{j})}=\bigcup_{m=0}^{f-1}D_{l+mp^{j-1}}^{(p^{j})}.$$
\end{remark}
\begin{remark}
 Euler quotient has been used to construct sequence with high linear complexity. In \cite{Chen3}, Chen et al. proposed a class of
 $p^{2}$-periodic binary sequences derived from the Euler quotients and investigated
 the $k$-error linear complexity, under the assumption of 2 being a primitive root modulo $p^{2}$ in \cite{Chen5}. Inspired by this construction,
  we proposed a general construction of binary sequences based on Euler quotient (and specially, Fermat quotient) with
  flexible support sets in \cite{Ye,Ye2}. Then, Du et al. defined a class of d-ary sequence using the Euler quotient,
which can be regarded as a generalization of the binary case, and then analyze the linear complexity of the proposed sequence
\cite{Du}. For more details, please refer to above mentioned references and the references therein.
\end{remark}

\subsection{Auxiliary lemmas}

\begin{lemma}\

\begin{enumerate}

\item  [(1)] For any $n\geq 1$ and $0\leq l\leq p^{n}f-1$, if $u \bmod p^{n+1}\in D_{l'}^{(p^{n+1})}$ for some $0\leq l^{'}\leq p^{n}f-1$, we have
$$uD_{l}^{(p^{n+1})}=\{uv \bmod p^{n+1}:v\in D_{l}^{(p^{n+1})}\}=D_{\langle l+l'\rangle_{p^{n}f}}^{(p^{n+1})},$$
where $\langle i\rangle_{e}$ denotes the least nonnegative integer congruent to
$i$ modulo $e$ for an integer $i$ and a positive integer $e$.

\item  [(2)] For any $1\leq n\leq n^{'}$, $0\leq l\leq p^{n^{'}}-1$ and $0\leq m\leq f-1$, we have
$$D_{(l,m)}^{(p^{n'+1})} \bmod p^{n+1}=D_{(\langle l\rangle_{p^{n}},\langle \lfloor\frac{l}{p^{n}}\rfloor+m\rangle_{f})}^{(p^{n+1})}.$$
Specially,
$$D_{(l,m)}^{(p^n)} \bmod p=D_{(0,\langle l+m\rangle_{f})}^{(p)}.$$
\end{enumerate}
\end{lemma}
Proof: Since
$$D_{l}^{(p^{n+1})}=\{g^{l+kfp^{n}}\bmod p^{n+1}:0\leq k <e-1\},$$
we have
$$uD_{l}^{(p^{n+1})}\bmod p^{n+1}=\{g^{l+l'+(k+k')fp^{n}}\bmod p^{n+1}:0\leq k <e-1\},$$
for an integer $k'\in \mathbb{Z}_{e}$. Then we get the property $(1)$. Similarly,
$$D_{(l,m)}^{(p^{n'+1})}\bmod p^{n+1}=\{g^{l+(kf+m)p^{n'}}\bmod p^{n+1}:0\leq k <e-1\},$$
where $l\in \mathbb{Z}_{p^{n'}},~f\in \mathbb{Z}_{f}$. The exponent is computed modula $p^{n}(p-1)$, the result then follows.
\hfill$\Box$

Let $D_{(l,m)}^{(p^n)}(x)=\displaystyle \sum_{u \in D_{(l,m)}^{(p^n)}}x^{u}$ , for $n\ge 1$ and $\theta_{n}$ be a primitive $p^{n}$-th root
of unity in an extension field of $\mathbb {F}_2$.
\begin{lemma}\
For any $0\leq l\leq p^{n-1}-1$, if $a,a'\in D_{l}^{(p^n)}$ we have
$$D_{l}^{(p^n)}(\theta_{n}^{a})=D_{l}^{(p^n)}(\theta_{n}^{a'}).$$

\end{lemma}
Proof: It is easy to check by Lemma 1 and the definition of $D_{(l,m)}^{(p^n)}(x)$, so we omit the proof.
\hfill$\Box$

\begin{lemma}\
Let symbols be the same as before. If
$$2^{p-1} \not\equiv 1 \bmod p^2 \ \mbox{and} \ \ \gcd\big(\frac{p-1}{{\rm{ord}}_p(2)},f\big)=1,$$
then for any integer
$n,~0<n\leq r$,
$$S(\theta_{r}^{u})=0, \mbox{~for~all}~u\in p^{r-n}\mathbb{Z}_{p^{n}}^{*} ,$$
if and only if
there exist $u\in p^{r-n}\mathbb{Z}_{p^{n}}^{*}$, such that $S(\theta_{r}^{u})=0$.
\end{lemma}
Proof: Assume that  $S(\theta_{r}^{u})=0$ for a $u\in p^{r-n}D^{(p^{n})}_{l}$, then we have $S(\theta_{r}^{u})=0$ for $all~u\in p^{r-n}D^{(p^{n})}_{l}$ by Lemma 2. Since $2^{p-1} \not\equiv 1 \bmod p^2$, we know $2\notin D_u^{(p^2)}$ with $p\mid u$. By Lemma 1-(1), we have  $2\in D_{u_{j}}^{(p^j)}$ with $\gcd(u_{j},p)=1$ for any $j\geq2$.
Denote $b={\rm{ord}}_p(2)$. Then by the definition of $D_{u_{1}}^{(p)}$, we have $u_{1}\equiv\frac{p-1}{b}d\bmod f$, where $\gcd(d, p-1)=1$. Therefore, $\gcd(u_{j},f)=\gcd(u_{1},f)=1$ since $\gcd(\frac{p-1}{b},f)=1$.
 Note that $2^aD^{(p^{n})}_{l}=D^{(p^{n})}_{l+au_{n}}$ for any integer $a$. Hence, it will run through $\mathbb{Z}_{p^{n}}^{*}$ when $a$ runs through $fp^{n-1}$. Since $0=S(\theta_{r}^{u})^{2^a}=S(\theta_{r}^{2^au})$ over $\mathbb {F}_2$, the result then follows.
\hfill$\Box$
\\


\begin{lemma}\
For any given integer $v$ in $D_{(l,m)}^{(p^{n+1})}$ and $T_{(v,n)}=\{v,v+p^n,v+2p^n,\ldots,v+(p-1)p^n\}$, we have
$$T_{(v,n)}\bigcap D_{(l+ip^{n-1},\langle m-i\rangle_{f})}^{(p^{n+1})}=1,~for~i=0,1,\ldots,p-1.$$
Furthermore, $v$ could be selected randomly from $D_{(l+ip^{n-1},\langle m-i\rangle_{f})}^{(p^{n+1})},~for~i=0,1,\ldots,p-1.$
\end{lemma}
Proof: By the properties of Euler quotient, we know that $Q_{p^n}(v+kp^n)$ runs through $l+ip^{n-1}$ for $i=0,1,\ldots,p-1$, when $k$ run through $i=0,1,\ldots,p-1$. Therefore, we have
$$T_{(v,n)}\bigcap \widetilde{D}_{l+ip^{n-1}}^{(p^{n+1})}=1,~for~i=0,1,\ldots,p-1.$$
In addition, let $T_{(v,n)}\bigcap \widetilde{D}_{l+ip^{n-1}}^{(p^{n+1})}\in D_{(l+ip^{n-1},m_{i})}^{(p^{n+1})}$, we have
 $$D_{(l+ip^{n-1},m_{i})}^{(p^{n+1})}=D_{(0,\langle l+m_{0}\rangle_{f})}^{(p)}\bmod p~,$$
since
$$v\equiv v+p^n \equiv \cdots \equiv v+(p-1)p^n \bmod p~.$$
By Lemma 1, we have
$$l+ip^{n-1}+m_{i}\equiv l+m_{0}\bmod f,$$
that is,
$$m_{i}+i\equiv m_{0}\bmod f~.$$
The result then follows.
\hfill$\Box$

\begin{lemma}\
Let the symbol be the same as before, there exists $Y\subseteq \mathbb{Z}_{p^{n}}^*$ such that
$$\bigcup_{v\in Y}T_{(v,n)}=\bigcup_{(i,j)\in X_{n+1}} D_{(i,j)}^{(p^{n+1})},$$
if and only if
$$X_{n+1}=\bigcup_{(l,m)\in X}\{(i,j):(i,j)=(l+kp^{n-1},\langle m-k\rangle_{f}),k=0,1,\ldots,p-1\}~,$$
where $X$ is a subset of $\mathbb{Z}_{p^{n-1}}\times \mathbb{Z}_{f}$.
\end{lemma}

Proof: By Lemma 4, if $v\in D_{(l,m)}^{(p^{n+1})}$ we have
$$T_{(v,n)}\subseteq \bigcup_{i=0}^{p-1} D_{(l+ip^{n-1},\langle m-i\rangle_{f})}^{(p^{n+1})}.$$
Hence,
$$\bigcup_{v\in D_{(l,m)}^{(p^{n+1})}}T_{(v,n)}\subseteq \bigcup_{i=0}^{p-1} D_{(l+ip^{n-1},\langle m-i\rangle_{f})}^{(p^{n+1})}.$$
Note that $T_{(v,n)}\bigcap T_{(v',n)}=\emptyset~for~any~v\neq v'\in D_{(l,m)}^{(p^{n+1})}$ and the size of two sets on both sides of above equation is equal to $ep$, so
$$\bigcup_{v\in D_{(l,m)}^{(p^{n+1})}}T_{(v,n)}= \bigcup_{i=0}^{p-1} D_{(l+ip^{n-1},\langle m-i\rangle_{f})}^{(p^{n+1})}.$$
Since at least one element of $T_{(v,n)}$ belongs to $\mathbb{Z}_{p^{n}}^*$, we can choose the $v$ from $\mathbb{Z}_{p^{n}}^*$ to construct $T_{(v,n)}$. Therefore, there exist $U\subseteq \mathbb{Z}_{p^{n}}^*$ such that
$$\bigcup_{v\in U}T_{(v,n)}= \bigcup_{i=0}^{p-1} D_{(l+ip^{n-1},\langle m-i\rangle_{f})}^{(p^{n+1})}.$$
Based on $T_{(v,n)}\bigcap T_{(v',n)}=\emptyset$ for any $v\neq v'\in \mathbb{Z}_{p^{n}}^*$ and $D_{(i,j)}^{(p^{n+1})} \bigcap D_{(i',j')}^{(p^{n+1})}$ for any $(i,j)\neq (i',j')\in \mathbb{Z}_{p^{n}}\times \mathbb{Z}_{f}$, the set $\bigcup_{(i,j)\in X_{n+1}} D_{(i,j)}^{(p^{n+1})}$ should be the union of some set which could be form as $\bigcup_{i=0}^{p-1} D_{(l+ip^{n-1},\langle m-i\rangle_{f})}^{(p^{n+1})}$.
The result then follows.
\hfill$\Box$

\subsection{An important lemma}
  To determine the exact values of $\delta_{i}$, $i=0,\ldots,r-1$ in the main theorem, we will introduce an important and useful lemma in this section. Before doing this, let us introduce a special enumeration method firstly. In detail, we will consider the representation of $C_{1}$ modula $p^n$ for different $n$. Hence, some elements may be repeated. Since we consider the sequence over $\mathbb{F}_2$, the occurrence number of each element in $C_{1}$ is required to be  modulo $2$. For example, $\{1,3,5\}\bmod 4=\{3\}$ and $\{1,3,5\}\bmod 2=\{1\}$. Due to $T_{(v,n)}=T_{(v+kp^n,n)} \bmod p^{n+1}$ for any integer $k$, we denote $T_{(v,n)}$ only for $v\in \mathbb{Z}_{p^{n}}$ here and hereafter.

\begin{lemma}\
Let symbols be the same as before. If
$$2^{p-1} \not\equiv 1 \bmod p^2 \ \ \mbox{and} \ \ \gcd(\frac{p-1}{{\rm ord}_{p}(2)},f)=1,$$ then for any
$n,~0\leq n\leq r-1$,
$$S(\theta_{r}^{u})=0, \mbox{~for~all}~u\in p^{r-n-1}\mathbb{Z}_{p^{n+1}}^{*} ,$$
if and only if one of the following conditions achieved
 \begin{itemize}
  \item [(i)] For $j<n$, there exists a set $W\subseteq\mathbb{Z}_{fp^{n-j-1}}$, such that
  $$X_{r-j}\bmod fp^{n-j}=\bigcup_{w\in W}\{w+kfp^{n-j-1}:k=0,1,\ldots,p-1\},$$
 and
$$\bigcup_{(l,m)\in Y_{r-n}} p^{n}D^{(r-n)}_{(l,m)} \bmod p^{n+1}=T_{(0,n)}\setminus\{0\},$$
   $$1+\sum_{j> n}e|X_{r-j}|\equiv 1\bmod 2.$$

\item [(ii)]For $j<n$, there exists a set $W\subseteq\mathbb{Z}_{fp^{n-j-1}}$, such that
  $$X_{r-j}\bmod fp^{n-j}=\bigcup_{w\in W}\{w+kfp^{n-j-1}:k=0,1,\ldots,p-1\},$$
 and
$$\bigcup_{(l,m)\in Y_{r-n}} p^{n}D^{(r-n)}_{(l,m)} \bmod p^{n+1}=\emptyset,$$
   $$1+\sum_{j> n}e|X_{r-j}|\equiv 0\bmod 2.$$
\end{itemize}

\end{lemma}

Proof:
Assume that
$S(\theta_{r}^{u})=0, \mbox{~for~all}~u\in p^{r-n_{0}-1}\mathbb{Z}_{p^{n_{0}+1}}^{*} ,$ for a fixed integer $n_0$ , where
$0\leq n_0\leq r-1$.
Since all ($\phi(p^{n_{0}+1})$ many) elements $\theta_{r}^{u}$ for $u\in p^{r-n_{0}-1}\mathbb{Z}_{p^{n_{0}+1}}^{*}$ are all roots of
$\Phi_{p^{n_{0}}}(x)=1+x^{p^{n_{0}}}+x^{2p^{n_{0}}}+\cdots+x^{(p-1)p^{n_{0}}}\in \mathbb{F}_{2}[x]$, we have
$\Phi_{p^{n_{0}}}(x)\mid S(x)$ in an extension field of $\mathbb{F}_{2}$. Then there exists a polynomial $\Psi(x)$
over $\mathbb{F}_{2}$ such that
\begin{equation}{\label{def-1}}
S(x)\equiv\Phi_{p^{n_{0}}}(x)\Psi(x)(\bmod ~x^{p^{n_{0}+1}}-1).
\end{equation}
Note that
$$x^{p^{n_{0}}}\Phi_{p^{n_{0}}}(x)\equiv\Phi_{p^{n_{0}}}(x)(\bmod~ x^{p^{n_{0}+1}}-1).$$
We can restrict $\deg(\Psi(x))<p^{n_{0}}$. Since the exponents of $\Phi_{p^{n_{0}}}(x)\Psi(x)$ can be form as $\bigcup_{v\in U}T_{(v,n_{0})}~~\mbox{for~some}~U\subseteq\mathbb{Z}_{p^{n_{0}}}$. Therefore, the equation (3) holds if and only if there exist $U\subset\mathbb{Z}_{p^{n_{0}}}$, such that
\begin{equation}{\label{def-2}}
C_{1} \bmod p^{n_{0}+1}=\bigcup_{v\in U}T_{(v,n_{0})}.
\end{equation}
Due to $$C_{1}=\displaystyle\bigcup_{j=1}^{r}  \bigcup_{i\in X_{j}}p^{r-j}D_{i}^{(p^j)}\bigcup\{0\},$$
we consider the case of $j\leq n_{0}$ firstly. Note that the elements in $p^{j}D^{(p^{r-j})}_{(l,m)}\bmod p^{n_{0}+1}$ only appear in the $T_{(v,n_{0})}$ if $v\in p^j\mathbb{Z}_{p^{n_{0}-j}}^*$.
Therefore, we have
\begin{equation}{\label{con}}
X_{r-j}\bmod fp^{n_{0}-j}=\bigcup_{(l,m)\in X}\{l+kp^{n_{0}-j-1}+\langle m-k\rangle_{f}p^{n_{0}-j},k=0,1,\ldots,p-1\}~,
\end{equation}
where $X$ is a subset of $\mathbb{Z}_{p^{n_{0}-j-1}}\times \mathbb{Z}_{f}$ based on Lemma 5. Note that the difference of adjacent element in the right set from above equation is $p^{n_{0}-j}-p^{n_{0}-j-1}\bmod p^{n_{0}-j}f$. Therefore, the difference will run through $kefp^{n_{0}-j-1}\bmod p^{n_{0}-j}f$ for $k=0,1,\ldots,p-1$. Hence, we get
$$X_{r-j}\bmod fp^{n_{0}-j}=\bigcup_{w\in W}\{w+kfp^{n_{0}-j-1}:k=0,1,\ldots,p-1\},$$
where $W$ is a subset of $\mathbb{Z}_{fp^{n_{0}-j-1}}$.
In particular, the aforementioned condition is impossible to achieved when $j=n_{0}$. The nearest
case is
\begin{equation}{\label{con1}}
\bigcup_{(l,m)\in Y_{r-n_{0}}} p^{n_{0}}D^{(r-n_{0})}_{(l,m)} \bmod p^{n_{0}+1}=T_{(0,n_{0})}\setminus\{0\}.
\end{equation}
If $j> n_{0}$,
$p^{j}D^{(p^{r-j})}_{(l,m)}\bmod p^{n_{0}+1}$ equals to $\{0\}~ {\rm or} ~\emptyset$. Hence,
$$1+\sum_{j> n_{0}}e|X_{r-j}|\equiv 1\bmod 2$$ and (\ref {con1}) will ensure that the equality (\ref{con}) also holds in the case $j=n_{0}$.
Another situation for the case $j=n_{0}$ is
$$\bigcup_{(l,m)\in Y_{r-n_{0}}} p^{n_{0}}D^{(r-n_{0})}_{(l,m)} \bmod p^{n_{0}+1}=\emptyset,$$
and $1+\sum_{j> n_{0}}e|X_{r-j}|\equiv 0\bmod 2$.
Combing above cases, we complete the proof.
\hfill$\Box$

\subsection{Proof of the main theorem}
Now we are ready to give a proof of the main theorem.

{\bf Proof of the main theorem.}
According to Lemma 3, the linear complexity of the sequence $S$ defined as (2) should be the form as we represent in the main theorem.
By using Lemma 6 repeatedly, we can get the exact value of $\delta_{i}$. That is, if the condition of Lemma 6 is achieved, then $\delta_{i}=0$. Otherwise,  $\delta_{i}=1$. The value of $\delta$ depends on $S(1)$, i.e. the weight of the sequence. The result then follows.
\hfill$\Box$

\begin{remark}
If the condition in Lemma 6 is not satisfied for $n=r$,
then the linear complexity of the sequence defined in $\rm (2)$ is at least $(p-1)p^{r-1}$, which is larger than half of its period. In particular, if $X_{1}\neq \emptyset~or~\mathbb{Z}_{f}\setminus\{0\}$,  the linear complexity of the sequence is at least larger than
 $(p-1)p^{r-1}$.
\end{remark}

\begin{remark}
It is known that the primes satisfying the condition of $2^{p-1} \equiv 1 \bmod p^2$
are very rare. Up to the present, there are only two such primes 1093 and
3511, up to $6.7\times 10^{15}$ \cite{FG}. If 2 is a primitive root of prime $p$, then the condition in Lemma 3 will be achieved.
By Artin's conjecture on primitive roots, the probability of 2 being a primitive root of prime $p$ is at least $37\%$. Therefore, the assumption in our theorem is acceptable.
\end{remark}

\subsection{Examples}
By repeatedly using Lemma 6, we can get the linear complexity of sequence defined in (2). Here, we give two examples to illustrate  the efficiency of our method, both of which have been verified by using the Magama.
\begin{example}
Let $p=11,~r=2,~e=2,~f=5,$\\
$$X_{1}=\{0,1,2,3,4\},~X_{2}=\{0,1,\ldots,24\}.$$ Then
$$X_{2}\bmod f=\{0,1,2,3,4\}$$
Then
condition in Lemma 6 is achieved when modulo $p$. Hence,
$$L=1+(p-1)p^{2}=111,$$
and the balanced sequence $(s_{i})$ defined in (2) is
\begin{equation*}
\begin{split}
&[ 1, 1, 1, 0, 1, 1, 0, 1, 1, 1, 1, 1, 0, 0, 1, 0, 1, 0, 1, 0, 1, 0, 1, 1, 0, 0,\\
&0, 0, 1, 1, 0, 0, 1, 1, 0, 0, 1, 0, 0, 1, 1, 1, 0, 0, 1, 0, 1, 0, 0, 1, 0, 0, 0,\\
&0, 0, 1, 1, 1, 1, 0, 0, 0, 0, 1, 1, 1, 1, 0, 0, 0, 0, 0, 1, 0, 0, 1, 0, 1, 0, 0,\\
&1, 1, 1, 0, 0, 1, 0, 0, 1, 1, 0, 0, 1, 1, 0, 0, 0, 0, 1, 1, 0, 1, 0, 1, 0, 1, 0,\\
&1, 0, 0, 1, 1, 1, 1, 1, 0, 1, 1, 0, 1, 1 ].
\end{split}
\end{equation*}
The linear complexity of this sequence is 111.
\end{example}

\begin{example}
Let $p=5,~r=3,~e=2,~f=2,$\\
\begin{equation*}
\begin{split}
X_{1}=\{0,1\},~X_{2}=\{0,2,4,6,8\},~X_{3}=\{0,10,20,30,40\}.
\end{split}
\end{equation*}
 Then
condition in Lemma 6 is achieved when modulo $p^3$.
Hence,
$$L=1+(p-1)(1+p)=25,$$
and the sequence $(s_{i})$ defined in (2) is
\begin{equation*}
\begin{split}
&[ 1, 1, 0, 0, 0, 1, 0, 0, 0, 0, 0, 0, 0, 0, 0, 0, 0, 0, 0, 0, 1, 0, 0, 0, 1, 1,\\
&1, 0, 0, 0, 1, 0, 0, 0, 0, 0, 0, 0, 0, 0, 0, 0, 0, 0, 0, 1, 0, 0, 0, 1, 1, 1, 0,\\
&0, 0, 1, 0, 0, 0, 0, 0, 0, 0, 0, 0, 0, 0, 0, 0, 0, 1, 0, 0, 0, 1, 1, 1, 0, 0, 0,\\
&1, 0, 0, 0, 0, 0, 0, 0, 0, 0, 0, 0, 0, 0, 0, 1, 0, 0, 0, 1, 1, 1, 0, 0, 0, 1, 0,\\
&0, 0, 0, 0, 0, 0, 0, 0, 0, 0, 0, 0, 0, 1, 0, 0, 0, 1 ].
\end{split}
\end{equation*}
The linear complexity of this sequence is 25.
\end{example}

\section{Conclusion}

In this paper, we present a generic construction of binary sequences with period $p^r$, which can be regarded as a generalization of the construction introduced by Xiao et al. By using the Euler quotient, we determined the linear complexity of the proposed sequence over $\mathbb{F}_2$ under the assumption of $2^{p-1} \not\equiv 1 \bmod p^2$ and $\gcd(\frac{p-1}{{\rm{ord}}_{p}(2)},f)=1$. As a byproduct, we showed that a conjecture proposed by Xiao et al. is correct under the aforementioned assumption. It will be interesting to determine the exact value of linear complexity in the case of $2^{p-1} \not\equiv 1 \bmod p^2$ without the condition $\gcd(\frac{p-1}{{\rm ord}_{p}(2)},f)=1$.

\section*{Acknowledgements}
This work was supported by National Natural Science
Foundation of China (No. 61772292, 61772476), Foundation of Fujian Educational Committee (No. JAT170627), Fujian Normal University Innovative
Research Team (No. IRTL1207).

\bibliographystyle{latex8}
\bibliography{latex8}


\end{document}